\documentclass[aps,nofootinbib,twocolumn,prd,eqsecnum,showpacs,showkeys,preprintnumbers,nofootinbib]{revtex4-1}
\usepackage{graphicx}
\usepackage{amsmath}
\usepackage{amsfonts}
\usepackage{amssymb}
\usepackage{color}
\usepackage{mathrsfs}
\usepackage{epstopdf}
\usepackage{url}
\usepackage{footnote}
\usepackage{textcomp}
\usepackage{hyperref}
\usepackage{dcolumn}
\usepackage[caption=false]{subfig}
\usepackage{cases}
\usepackage{bm}

\begin{document}

\title{Smoking guns of a bounce in modified theories of gravity through the spectrum of the gravitational waves.}

\author{Mariam Bouhmadi-L\'{o}pez $^{1,2,3,4}$}
\email{mariam.bouhmadi@ehu.es}
\author{Jo\~{a}o Morais $^{2}$}
\email{joao.morais@ist.utl.pt}
\author{Alfredo B. Henriques $^{2}$}
\email{alfredo.henriques@fisica.ist.utl.pt}
\date{\today}

\affiliation{
${}^1$ Instituto de Estructura de la Materia, IEM-CSIC, Serrano 121, 28006 Madrid, Spain\\
${}^2$ Centro Multidisciplinar de Astrof\'{\i}sica - CENTRA, Departamento de F\'{\i}sica, Instituto Superior T\'ecnico, Av. Rovisco Pais 1,1049-001 Lisboa, Portugal\\
${}^3$ Department of Theoretical Physics, University of the Basque Country
UPV/EHU, P.O. Box 644, 48080 Bilbao, Spain\\
${}^4$ IKERBASQUE, Basque Foundation for Science, 48011, Bilbao, Spain\\}

\date{\today}

\begin{abstract}
We present an inflationary model preceded by a bounce in a metric theory a l\'{a} $f(R)$ where $R$ is the scalar curvature of the space-time. The model is asymptotically de Sitter such that the gravitational action tends asymptotically to an Einstein-Hilbert action with an effective cosmological constant, therefore modified gravity affects only the early stages of the universe. We then analyse the spectrum of the gravitational waves through the method of the Bogoliubov coefficients by two means: taking into account the gravitational perturbations due to the modified gravitational action in the $f(R)$ setup and by simply considering those perturbations inherent to the standard Einstein-Hilbert action. We show that there are distinctive (oscillatory) signals on the spectrum for very low frequencies; i.e. corresponding to modes that are currently entering the horizon.
\end{abstract}


\maketitle

\section{\label{sec:Intro}Introduction}

The inflationary paradigm is the most accepted and successful theoretical framework to explain the homogeneity, flatness and horizon problems we observe today. However, inflation does not give any clues as to what mechanism is responsible for the origin of the inflationary era of the universe \cite{LiddleLyth}. As such, many different model have been developed, from scalar fields to modified theories of gravity that try to describe the early accelerated expansion of the universe that leads to the seeds of the universe we observe today (cf. for example \cite{Lidsey:1999mc,Bassett:2005xm,Langlois:2010xc,Maartens:2010ar}). In order to gain some insight on the dynamics of the early universe and to begin to discriminate between these models, one needs to probe the spectrum of the cosmological perturbations: for example the scalar or tensor perturbations, that arise from the quantum vacuum fluctuations during the inflationary period. In fact, since the evolution of these perturbations depends on the characteristics of the mechanism causing the acceleration of the universe, it is reasonable to assume that a given inflationary model will have some characteristic imprints on the energy spectrum of the gravitational waves (GWs).

Another issue is that the inflationary paradigm does not solve the big bang singularity; indeed it is expected that such a singularity would be resolved within the paradigm of quantum cosmology \cite{claus}. A plausible candidate regarding this is loop quantum cosmology (LQC), which predicts a bounce instead of the big bang singularity \cite{bojowald}. Bouncing cosmologies have a long history and can appear for example in modified theories of gravity {\color{black}\cite{Novello1,Carloni1,Barragan1,Cai:2011tc,Biswas:2005qr}} and lead to the hypothesis of cyclic universes \cite{Piao:2005ag,Cai:2009in}.

In this paper, we propose a bouncing scenario within the context of modified theories in the metric formalism, more precisely we will consider an $f(R)$ model. These kind of models have been quite popular recently as they could describe the current acceleration of the universe without invoking a dark energy component, i.e. the modification in the modified Friedmann equation mimics the effects of a dark energy component with the advantage that the acceleration is a pure gravitational effect and is not due to an unknown dark energy component (see for example Refs.~\cite{Sotiriou:2008rp,Olmo:2011uz} for reviews on $f(R)$ gravity and Refs.~\cite{Nojiri:2006ri,Capozziello2} for reviews on extended theories of gravity). Despite that the $f(R)$ modifications can cause some conflict with observational constrains, (e.g. solar system tests \cite{Berry:2011pb}), a proper chameleonic mechanism was suggested in Refs.~\cite{Hu:2007nk,Starobinsky:2007hu}.

In addition, one of the pioneer inflationary models was formulated within the context of modified gravity and was due to Starobinsky \cite{starobinsky}. In this model the modification of gravity was a result of quantum fluctuations of fields and gravity itself.

The authors of Ref.~\cite{Olmo:2008nf} map the modified Friedmann equation in LQC to that of Palatini $f(R)$ theory, \cite{Olmo:2011uz}. More specifically, they found the right $f(R)$ function that has to be considered to obtain a bouncing cosmology of the kind of LQC, providing for the first time the covariance of the theory through an effective action. Our approach to a bouncing cosmology is different from the one proposed in~\cite{Olmo:2008nf} for the two following reasons: first we assume a metric formalism, second we choose a specific scale factor evolution. We essentially prefer the metric formalism to the Palatini one in the present work because the cosmological perturbations are easier to handle on the metric formalism as compared with the Palatini one. Moreover, on the Palatini $f(R)$ formalism non-vacuum static spherically symmetric objects exhibit curvature singularities, casting doubt on whether Palatini $f(R)$ gravity can be considered as giving viable alternatives to GR as suggested in \cite{Barausse:2007pn}.

The bounce we will consider in our model is followed by an inflationary era which is asymptotically de Sitter where, in addition, the gravitational action approaches the Einstein-Hilbert action with an effective cosmological constant on that regime, such that the modification to Einstein's General Relativity (GR) affects exclusively the very early universe, around the bounce and a few e-folds after that. We will constrain the model obtaining the spectrum of the stochastic gravitational fossil as would be measured today. Such an analysis will be carried out using a generalization for $f(R)$-gravity of the method of the continuous Bogoliubov coefficients first introduced in \cite{Parker1,Starobinsky1,Allen:1987bk}, and later applied in \cite{Henriques1, Moorhouse1, Mendes1}. 
One of these Bogoliubov coefficients gives the density of gravitons of the universe. An alternative approach to obtain the spectrum of the GWs in modified theories of gravity was developed in \cite{Mukhanov:1990me,Capozziello:2007zza,Capozziello:2007vd,Capozziello:2008fn}

The outline of the paper is as follows. In section \ref{sec:fR}, we define the behaviour of the bounce through its scale factor and derive the appropriate $f(R)$ action compatible with such an evolution of the early universe (see Refs.~\cite{Dunsby:2010wg,Carloni:2010ph} for works of reconstruction methods in $f(R)$ gravity). In section \ref{sec:Spectrum}, we
summarise the methodology used to obtain the spectrum of the stochastic GWs which is based on the Bogoliubov coefficients. We present in section \ref{sec:Numerical} the spectrum of the GWs for the model introduced in section \ref{sec:fR} using both a GR setup and an  $f(R)$ setup to analyse the evolution of the gravitational perturbations. Finally, in section \ref{sec:Conclusions} we present our conclusions.

\section{\label{sec:fR}A bounce in \lowercase{\textit{f}}(R)-metric gravity}

In this paper we introduce a phenomenological bounce in the early universe as a way to avoid the big bang singularity. A bounce at time $t_b$ can be characterized in terms of the scale factor by:
\begin{equation}
	\label{eq: bounce}
	\dot{a}(t_b)=0 ~~~~ \text{and} ~~~~ \ddot{a}(t_b)>0.
\end{equation}

Inspired on the de Sitter solution for a closed Friedmann-Lema\^{\i}tre-Robertson-Walker (FLRW) space-time, we define the scale factor around the bounce as:
\begin{equation}
	\label{eq: scalefactor}
	a(t) = a_b \cosh\left[H_{\textrm{inf}}(t-t_b)\right]
\end{equation}
where $a(t)$ is the scale factor, $a_b$ is a constant quantifying the size of the universe at the bounce. In addition, $H_{\textrm{inf}}$ is related to the energy scale of inflation just after the bounce.\footnote{Notice at this regard that $\ddot{a}>0$ at all times and in particular for $t>t_b$. In addition, $\dot{a}>0$ for $t>t_b$.} For simplicity, we will set $t_b$ to be zero. 

While in GR the previous solution \eqref{eq: scalefactor} corresponds to a de Sitter space-time with spherical spatial section, here we are looking for a modified theory of gravity of the kind $f(R)$, such that Eq.~\eqref{eq: scalefactor} is a solution of the modified Friedmann equation of a FLRW universe with flat spatial section. This is a simple and straightforward way of modelling a bounce within the paradigm of modified theories of gravity. In addition, our choice is not only based on criterion of simplicity but also inspired on the fact that the above mentioned solution is asymptotically de Sitter in the past and the future (for space-times with flat spatial sections). Consequently, this procedure assures avoiding the initial singularity and guarantees a period of inflation leading to the first seeds of the structures we see nowadays. Also, the existence of such a bounce is prohibited for a flat space-time in GR unless exotic matter, which violates the null energy conditions, is invoked \cite{Molina1}. In this paper, we work within the framework of $f(R)$ theories of gravity which allow for a rich description of the universe (see for example \cite{Sotiriou:2008rp,Capozziello2}) and as we will show this approach allows for the kind of bounces described by Eq.~\eqref{eq: scalefactor}. For alternative approaches on bouncing cosmologies in modified theories of gravity cf. for example Refs.~\cite{Barragan1,Carloni1,Novello1}.

The action of an $f(R)$ gravity reads \cite{Capozziello2}:
\begin{equation}
	S = \frac{1}{2\kappa^2}\int f(R)\sqrt{-g}\textrm{d}^4x + S^{(m)},
\end{equation}
which leads to the modified Friedmann and Raychaudhury equations \cite{Capozziello2}:
\begin{equation}
	\label{eq: Fried1}
	H^2 = \frac{\kappa^2}{3f_R} \left( \rho - \frac{f(R)-f_RR}{2\kappa^2}-3H\frac{f_{RR}\dot{R}}{\kappa^2}\right),
\end{equation}
\begin{align}
	\label{eq: Fried2}
	2\dot{H} + 3H^2 &= \frac{\kappa^2}{f_R} \left( p + \frac{f(R)-f_RR}{2\kappa^2} \right.
	\nonumber\\
	&+ \left. \frac{f_{RRR}\dot{R}^2 +f_{RR}\ddot{R}+2Hf_{RR}\dot{R}}{\kappa^2}\right),
\end{align}
respectively. Here $H$ is the Hubble parameter, $\kappa^2=8\pi Gc^{-4}$, $G$ is the gravitational constant, $\rho$ and $p$ are the energy density and pressure of the matter content of the universe. The dot stands for a derivative with respect to the cosmic time and the $R$ subscript indicates a derivative with respect to the scalar curvature. We set the speed of light, $c$, to one on the present section and will set it back to its dimensional value on section~\ref{sec:Spectrum}, when we calculate the spectrum of the gravitational waves.

Using the above written equations (see Eq.~\eqref{eq: Fried1}), we can deduce the form of $f(R)$ such that the scale factor scales with the cosmic time as shown in Eq.~\eqref{eq: scalefactor}. More precisely, from Eq.~\eqref{eq: scalefactor} we derive the temporal evolution for the Hubble parameter, $H$, and the scalar curvature, $R$, in a spatially flat universe:
\begin{align}
	\label{eq: Hubble}
	H(t)= H_{\textrm{inf}}\tanh(H_{\textrm{inf}}t),
\end{align}
\begin{align}
	\label{eq: RicciScalar}
	R(t) = 6H_{\textrm{inf}}^2\left[1+\tanh^2(H_{\textrm{inf}}t)\right].
\end{align}

We will assume that the inflationary era is induced by modifications of gravity with respect to GR. This is a quite old idea first proposed by Starobinsky in its inflationary model \cite{starobinsky}. Therefore, we will assume that the energy density, $\rho$, and the pressure, $p$, are zero. Finally, by inserting Eqs.~\eqref{eq: scalefactor}, \eqref{eq: Hubble} and \eqref{eq: RicciScalar} in the modified Friedmann equation \eqref{eq: Fried1}, we obtain a constraint on $f(R)$; more precisely $f$ must satisfy a second order differential equation:
\begin{equation}
	\label{eq: diffeq_f}
	H\ddot{f}+(H^2 - 2\dot{H})\dot{f}+2H\dot{H}f =0.
\end{equation}

It is worthy to stress that this equation has two solutions \cite{Dunsby:2010wg,Carloni:2010ph}. Therefore, there is not a unique form of $f(R)$  given a specific evolution for the scale factor. We will have to impose some physical criteria to choose the appropriate one.
The solution of equation \eqref{eq: diffeq_f} can be written as a linear combination of the functions $f_1(t)$ and $f_2(t)$:
\begin{align}
	\label{eq: sol_f}
	f_1(t) = \sqrt{6\tanh^2(H_{\textrm{inf}}t)+3}\cos\left(\frac{\sqrt{3}}{2}\theta_1(t)+\theta_2(t)\right)
	\nonumber\\
	f_2(t) = \sqrt{6\tanh^2(H_{\textrm{inf}}t)+3}\sin\left(\frac{\sqrt{3}}{2}\theta_1(t)+\theta_2(t)\right),
\end{align}
where $\theta_1(t)$ and $\theta_2(t)$ are given by:
\begin{align}
	\label{eq: sol_theta}
	\theta_1(t) &= \text{sgn}(t)\arccos(1-2\tanh^2(H_{\textrm{inf}}t))
	\nonumber\\
	\theta_2(t) &= \arccos\left(\sqrt{\frac{\tanh^2(H_{\textrm{inf}}t)}{2\tanh^2(H_{\textrm{inf}}t)+1}}\right).
\end{align}
Here the sgn function is defined by $\textrm{sgn}(t)=2u(t)-1$ where $u(t)$ is the Heavyside step function \cite{abramowitz1}.

We will choose the linear combination of $f_1$ and $f_2$ such that the effective gravitational coupling $G^{\textrm{(eff)}}$ is always positive, or equivalently $f_R>0$ ($G^{\textrm{(eff)}}\equiv G/f_R$, as can be seen from \eqref{eq: Fried1}) . In order to impose this condition it is easier to rewrite $f_1$ and $f_2$ in terms of the scalar curvature \eqref{eq: RicciScalar}:
\begin{align}
	\label{eq: sol_f_R}
	f_1(R) &= \sqrt{R/H_{\textrm{inf}}^2-3}\cos\left(\frac{\sqrt{3}}{2}\theta_1(R)+\theta_2(R)\right),
	\nonumber\\
	f_2(t) &= \sqrt{R/H_{\textrm{inf}}^2-3}\sin\left(\frac{\sqrt{3}}{2}\theta_1(R)+\theta_2(R)\right).
\end{align}
Although we can invert Eq.~(2.7) analytically, for simplicity, we did not include the inverted relation between the cosmic time and the scalar curvature.

Therefore, the solution for $f(R)$ is given by:
\begin{equation}
	f(R)=C_1f_1(R)+C_2f_2(R).
\end{equation}

Differentiating \eqref{eq: sol_f_R} and imposing $f_R>0$  we obtain the following conditions for the linear coefficients $C_1$ and $C_2$:
\begin{equation}
	\label{eq: Acondition}
	C_1<0
	~~~~
	\textrm{and}
	~~~~
	\frac{C_2}{C_1} = \tan\left(\frac{\sqrt{3}}{2}\pi\right).
\end{equation}
We pick $C_1$ and $C_2$ to be of the form:
\begin{equation}
	C_1=C\cos\left(\frac{\sqrt{3}}{2}\pi\right)
	~~
	\textrm{and}
	~~
	C_2=C\sin\left(\frac{\sqrt{3}}{2}\pi\right),
\end{equation}
with $C$ a positive constant. It can be checked that this choice of coefficients satisfies Eq.~\eqref{eq: Acondition} and leads to the following expression for the $f(R)$ solution:
\begin{align}
	\label{eq: f(R)}
	f(R)& = C\sqrt{R/H_{\textrm{inf}}^2-3}
	\nonumber\\
	&\times\cos\left\{\frac{\sqrt{3}}{2}\left[\pi-\arccos\left(\frac{9-R/H_{\textrm{inf}}^2}{3}\right) \right]\right.
	\nonumber\\
	&- \left.\arccos\left(\sqrt{\frac{3}{2}\frac{R/H_{\textrm{inf}}^2-6}{R/H_{\textrm{inf}}^2-3}}\right) \right\}.
\end{align}

\begin{figure*}[!ht]
	\centering
	\subfloat[]{\includegraphics[width=.48\textwidth]{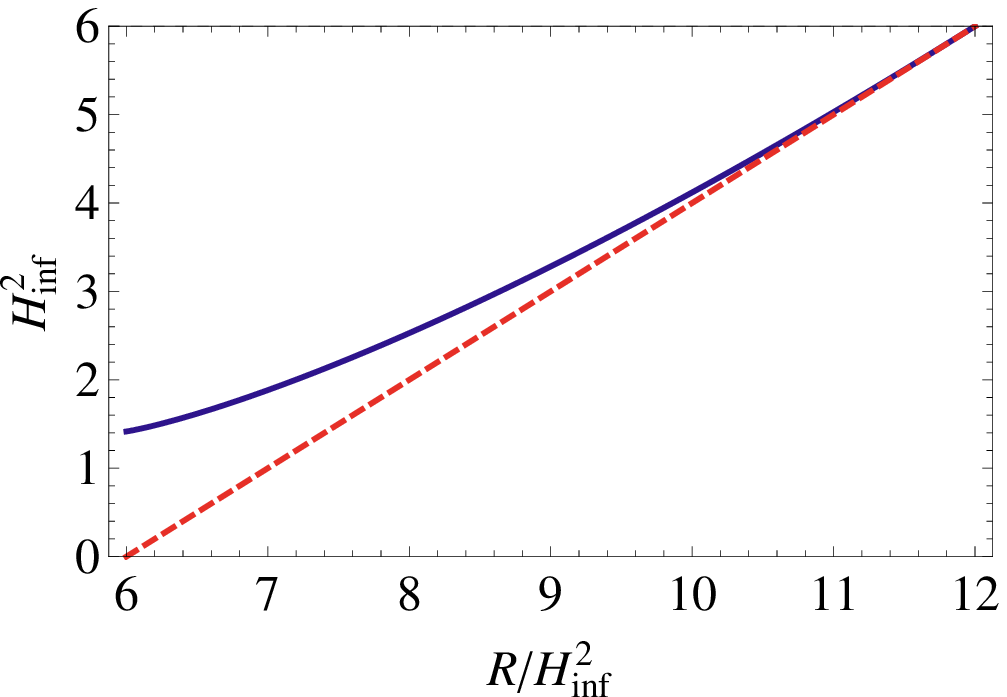}}
	\hfill
	\subfloat[]{\includegraphics[width=.48\textwidth]{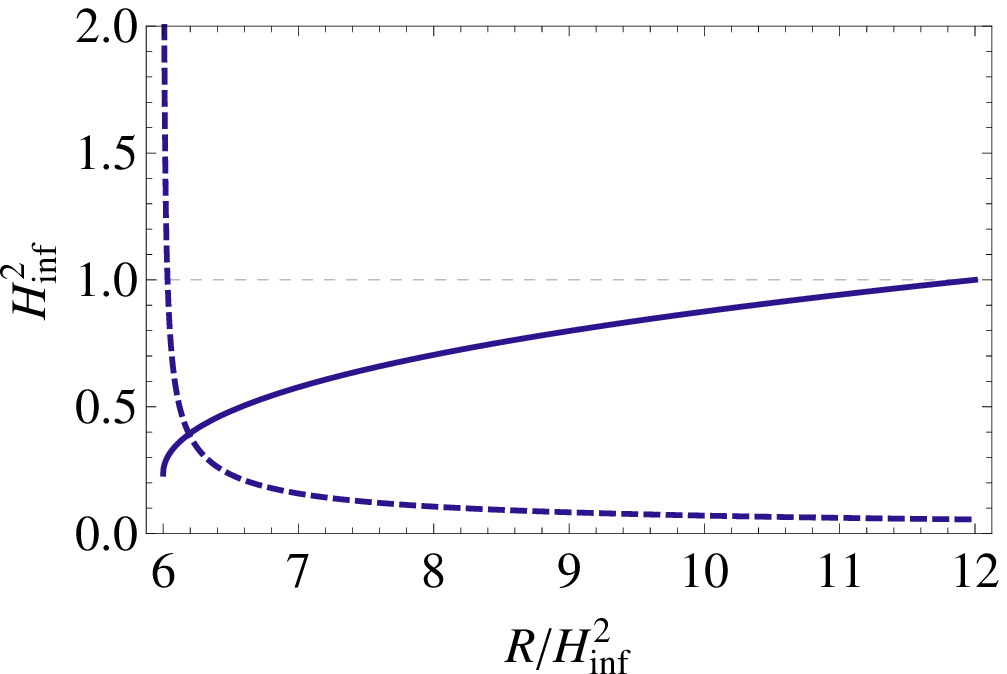}}\\
	\caption[$f(R)$]{\label{fig: f(R)}These plots show from left to right: (a) the behaviour of $f(R)$ as a function of $R/H_{\textrm{inf}}^2$ (see the continuous curve) and the Einstein-Hilbert action with a cosmological constant, $R-6H_{\textrm{inf}}^2$, (see the dashed curve) corresponding to the limiting behaviour of $f(R)$; (b) the behaviour of $f_R$ (see the continuous curve) and of $f_{RR}$ (see the dashed curve) as functions of $R/H_{\textrm{inf}}^2$.}
\end{figure*}

We are left with a single constant $C$ which can be fixed  by imposing that asymptotically, well inside the inflationary era:  (i) we recover the Einstein-Hilbert action with a cosmological constant, $R-2\Lambda$, and (ii) the gravitational coupling reduces to the gravitational constant $G$. Following this procedure we obtain $C=2H_{\textrm{inf}}^2$. In Fig.~\ref{fig: f(R)}, we plot the solution \eqref{eq: f(R)} and its first two derivatives with respect to the scalar curvature. Well inside the inflationary era, $R\approx12H_{\textrm{inf}}^2$, the second and higher derivatives of $f(R)$ become small and the action can be approximated as $f(R)\approx R-6H_{\textrm{inf}}^2$; we recover the standard behaviour of GR  with an effective cosmological constant $\Lambda\approx3H_{\textrm{inf}}^2$. For low values of the scalar curvature, $R$, the deviations from  GR become more significant, in particular close to the bounce.  In fact, at the bounce the second derivative and higher derivatives of $f(R)$ get very large. Despite the divergence of $f_{RR}$, and therefore of $f_{RRR}$, the Friedman and Raychandhuri equations, Eqs.~\eqref{eq: Fried1} and \eqref{eq: Fried2}, are well defined for all $R$. We will next show, however, that this behaviour does have serious implications especially in the scalar sector.

It is known that metric $f(R)$ gravity can be recast as a scalar-tensor theory, in particular a Brans-Dicke theory with Brans-Dicke:\footnote{Palatini $f(R)$ gravity is also a Brans-Dicke theory with Brans-Dicke parameter $\omega_0=3/2$\cite{Barragan1,Olmo:2005zr}.} parameter $\omega_0=0 $ and a new scalar degree of freedom, the scalaron, defined as $\phi=f_R$ \cite{Capozziello2,Olmo:2005zr,Olmo:2005hc,Faraoni3,Hwang1,Hwang2,Hwang3}. The squared mass of this field is given by \cite{Olmo:2005hc,Faraoni1}
\begin{equation}
	\label{eq: ScalMass} 
	m^2 = \frac{f_R - Rf_{RR}}{3f_{RR}}
\end{equation}
The squared mass of the scalaron in our case vanishes asymptotically, roughly when the scalar  curvature is about $R\approx8.7H_{\textrm{inf}}^2$. Unfortunately, it becomes negative for smaller values. This seems to imply that the model would be unstable on the scalar sector, and in particular the spectrum of the scalar perturbations might be ill defined.\footnote{It is usually assumed that the condition that the scalaron is not a ghost or a tachyon is closely related to the Dolgov-Kawasaki instability \cite{Faraoni2,Sotiriou:2008rp}, which implies $f_{RR}>0$. Notice that in our case despite having $f_{RR}>0$, the squared mass of the scalaron becomes negative around the bounce. The reason behind this behaviour is that close to the bounce the model deviates strongly from GR, while in deducing the condition $f_{RR}>0$ to avoid the  Dolgov-Kawasaki instability, it is assumed a small deviation from GR \cite{Faraoni2,Sotiriou:2008rp}.}


\section{\label{sec:Spectrum}Energy spectrum of gravitational waves}

We now analyse the GWs spectrum predicted by the model introduced in the previous section. In order to obtain the current energy spectrum of GWs \cite{Parker1,Starobinsky1,Sahni:1990tx}, we use the generalization for $f(R)$-gravity of the method of the continuous Bogoliubov coefficients introduced by Parker. Here a Bogoliubov transformation is applied to quantify the tensor perturbations of the metric in terms of time-fixed annihilation and creation operators and their linear coefficients $\alpha$ and $\beta$. These coefficients, which carry the time dependence of the vacuum state, fulfill the initial conditions $\alpha(t_{\textrm{ini}})=1$ and $\beta(t_{\textrm{ini}})=0$ at some time $t_{\textrm{ini}}$.\footnote{On Sec.~\ref{sec:Numerical}, we will set $t_{\textrm{ini}}$ before the bounce. Therefore, the value of the parameter $t_{\textrm{ini}}$ defines how much the universe contracts before the bounce by fixing the relation $a_{\textrm{ini}}/a_b$ between the initial value of the scale factor, $a_{\textrm{ini}}$, and the scale factor at the time of the bounce, $a_b$.} Furthermore, $|\beta(t)|^2$ gives the density of gravitons of the universe at a time $t$. A second linear transformation is introduced to describe the evolution of the graviton density by the more practical $X$ and $Y$ variables, which obey \cite{Henriques1,Mendes1,Moorhouse1,Sa1}:
\begin{equation}
		\label{eq: X'eq}
		X' = -ikY,
	\end{equation}
	\begin{equation}
		\label{eq: Y'eq}
		Y' = -\frac{i}{k}\left(k^2-\frac{a''}{a}\right)X.
	\end{equation}
Here a prime indicates a derivative with respect the conformal time, $\eta$ $(a=dt/d\eta)$. These new variables are related to the graviton density by $|\beta|^2 = |X-Y|^2/4$ and must respect the condition $\textrm{Re}(X\cdotp Y)=1$.
In a flat FLRW de Sitter space-time, the system of equations \eqref{eq: X'eq} and \eqref{eq: Y'eq} admits the solution \cite{Mendes1}:
\begin{equation}
	\label{eq: XdSsolutions}
	X(a)= \left(1+i\frac{aH}{k}\right)e^{i\frac{k}{aH}},
\end{equation}
\begin{equation}
	\label{eq: YdSsolutions}
	Y(a) = \left(1+i\frac{aH}{k} -\frac{a^2H^2}{k^2}\right)e^{i\frac{k}{aH}}.
\end{equation}

The graviton density, $|\beta|^2$, at the present time, $\eta_0$, defines the dimensionless relative logarithmic energy spectrum of GWs for the frequency $\omega$ as \cite{Allen:1987bk,Sa1}:
\begin{align}
\label{eq: Spectrum}
	\Omega_{\textrm{GW}} (\omega,\eta_0) = \frac{\hbar\kappa^2}{3\pi^2 c^5H^2(\eta_0)}\omega^4|\beta(\eta_0)|^2
\end{align}
where $c$ is the speed of light in vacuum and $\hbar$ is the reduced Planck constant. We set back all the units to obtain the spectrum of the gravitational waves.

In $f(R)$-gravity the same method can be applied by replacing $a''/a$ in Eq.~\eqref{eq: Y'eq} by $z''/z$ where $z=a\sqrt{f_R}$ \cite{Capozziello2,Mukhanov:1990me,Faraoni3,Hwang1,Hwang2,Hwang3}:
\begin{equation}
	\label{eq: Y'eqz}
	Y' = -\frac{i}{k}\left(k^2-\frac{z''}{z}\right)X.
\end{equation}
Using the definition of $z$ we can expand $z''/z$ into $a''/a$ plus a correction term due to the $f(R)$ modification, $\Xi$, that vanishes at the end of the $f(R)$ era:
\begin{align}
	\label{eq: zExp}
	\frac{z''}{z} &= \frac{a''}{a}+ \frac{a'}{a}\frac{f_R'}{f_R} + \frac{1}{2}\frac{f_R''}{f_R}-\frac{1}{4}\left(\frac{f_R'}{f_R}\right)^2
	\nonumber\\
	&= \frac{a''}{a} + \Xi.
\end{align}
Even though the definition of $\Xi$ contains terms that diverge at the bounce, because they involve $f_{RR}$ and higher derivatives, $\Xi$ is always finite \footnote{Notice that $f'_R$ and $f''_R$ can be written as linear combinations of $f_{RR}$ and $f_{RRR}$.}.

\begin{figure}[t]
	\includegraphics[width=\columnwidth]{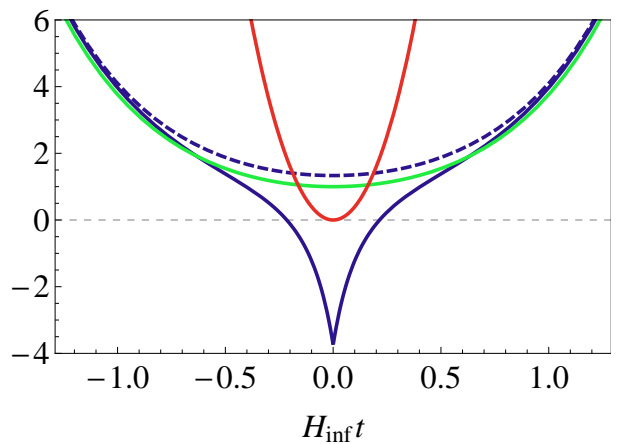}
	\caption[$z''/z$, $a''/a$ and $a^2H^2$ near the bounce]{\label{fig: Potencial_az}This Fig. shows: (i) the potential $z''/z$ (see the continuous blue curve) and its asymptotic behaviour (see the discontinuous blue curve); (ii) the potential $a''/a$ (see the green curve); (iii) the comoving wave-number $k_H^2=4\pi^2a^2H^2$ (see the red curve); as functions of cosmic time and near the bounce. All functions are  plotted in units of $a_b^2H_{\textrm{inf}}^2$.}
\end{figure}

The definition of the scale factor during the early universe, see Eq.~\eqref{eq: scalefactor}, as well as the solution obtained for $f(R)$, Eq.~\eqref{eq: f(R)}, allows us to cast both terms of Eq.~\eqref{eq: zExp} as functions of the cosmological time:
\begin{align}
	\label{eq: a''a1}
	\frac{a''}{a}(t) = H_{\textrm{inf}}^2a_b^2\left[2\cosh^2(H_{\textrm{inf}}t)-1\right],
\end{align}
\begin{align}
	\label{eq: fRcorrec}
	\Xi(t) =& H_{\textrm{inf}}^2a_b^2 \Biggl\{ 4-5\cosh^2(H_{\textrm{inf}}t) \Bigr.
	\nonumber\\
	&+ \left.\left[9\cosh^2(H_{\textrm{inf}}t)-6\right]Z(t)
	\right.\nonumber\\
	&- \left. \frac{\left[3\cosh^2(H_{\textrm{inf}}t)-2\right]^2}{4\sinh^2(H_{\textrm{inf}}t)}Z^2(t)\right\},
\end{align}
where the $Z(t)$ is defined as:
\begin{align}
	\label{eq: Z}
	Z(t) = \frac{\sqrt{3\sinh^2(H_{\textrm{inf}}t)}}{\cot\left(\frac{\sqrt{3}}{2}\theta_1(t)+\theta_2(t)\right)+\sqrt{3\sinh^2(H_{\textrm{inf}}t)}}.
\end{align}
The function $Z(t)$ is a monotonic increasing function of $|t|$ that goes from $Z(0)=0$ to the limiting value during the de Sitter-like expansion $Z(t\rightarrow\infty)=2/3$. In this regime $\Xi\approx H_{\textrm{inf}}^2a_b^2/3$, therefore, we can make the approximation:
\begin{equation}
	\label{eq: z''zLimit}
	\frac{z''}{z}(t\gg H_{\textrm{inf}}^{-1}) \approx H_{\textrm{inf}}^2a_b^2\left[2\cosh^2(H_{\textrm{inf}}t)-2/3\right].
\end{equation}
In Fig.~\ref{fig: Potencial_az}, we plot the behaviour of $a''/a$ and $z''/z$ near the bounce. The striking feature of the $f(R)$ modification is the fact that $z''/z$ becomes negative very close to the bounce $(z''/z<0$ approximately for $|t|<0.22H_{\textrm{inf}}^{-1})$. We note that, in GR, the potential $a''/a$ can become negative whenever $\rho-3p<0$, for example for stiff matter. The deviations from the GR term are maximum at the bounce: $\max|\Xi| =  |\Xi(0)|\approx4.7H_{\textrm{inf}}^2a_b^2$.

When the universe is well inside the inflationary era, the deviations of the action of our model from an Einstein-Hilbert action with a cosmological constant become small and equations \eqref{eq: Fried1} and \eqref{eq: Fried2} approach the form of the Friedmann and Raychaudhury equations derived from GR with an effective cosmological constant. We make use of this asymptotic behaviour by switching entirely to a GR description at this point, which allows us to describe the late time evolution of the universe by the $\Lambda$CDM model. To obtain a smooth transition between the $\Lambda$CDM model and the model introduced in Sec.~\ref{sec:fR}, we use a modified Generalized Chaplygin Gas (mGCG), see for example Ref.~\cite{Bouhmadi1} (for further works on the Chaplygin gas see Refs.~\cite{GonzalezDiaz:2002hr}). We choose to use a mGCG as it provides a simple phenomenological way to model a smooth transition from a de Sitter-like phase to the radiation dominated epoch. In this approach, we are essentially using two actions during the different stages of the evolution of the universe: the $f(R)$ action \eqref{eq: f(R)} is used until the later stages of inflation, where $R\in[6H_{\textrm{inf}}^2,12H_{\textrm{inf}}^2]$ and $a<a_1$ (see the paragraph just after Eq.\eqref{eq: z''z}); for ensuing times ($a>a_1)$, the Einstein-Hilbert action with a cosmological constant is employed. Given that (i) a time arrow is well defined from the bounce onward, (ii) the universe is asymptotically de Sitter around $a_1$ and that (iii) the $f(R)$ theory is linear on $R$ when it approaches $a_1$, we can connect the two actions continuously and evolve the tensor perturbations till the present time as we will show next.

The effective energy density in vacuum during the $f(R)$ era can be expressed as:
\begin{align}
	\label{eq: EffDensity}
	\rho^{\textrm{(eff)}} &= \frac{f_RR-f}{2\kappa^2 f_R} - \frac{3H}{\kappa^2}\frac{f_{RR}}{f_R}\dot{R}
	\nonumber\\
	&= \frac{3H^2}{\kappa^2} = \frac{3H_{\textrm{inf}}^2}{\kappa^2}\left[1-\left(\frac{a_b}{a}\right)^2\right].
\end{align}
Piecing it together with the densities for the mGCG and $\Lambda$CDM models, we can express the energy density of the universe throughout the different eras as follows:\footnote{To avoid confusion and ambiguity in the notation, we have adopted $\lambda$ for the parameter of the mGCG, rather than the usual symbol $\alpha$ which on the present paper denotes one of the Bogoliubov coefficients, thus differing from the notation in Ref.~\cite{Bouhmadi1}.}
\begin{equation}
	\label{eq: rhototal}
\rho=\begin{cases}
	\frac{3H_{\textrm{inf}}^2}{\kappa^2}\left[1-\left(\frac{a_b}{a}\right)^2\right] & \textrm{$f(R)$ era},\\
	\left[A+\frac{B}{a^{4(1+\lambda)}}\right]^{\frac{1}{1+\lambda}} & \textrm{mGCG era},\\
	\rho_r \left(\frac{a_0}{a}\right)^4 + \rho_m \left(\frac{a_0}{a}\right)^3 + 4\rho_{\Lambda} &\Lambda\textrm{CDM era}.
\end{cases}
\end{equation}
The constants $A$ and $B$ are fixed by comparing the asymptotic behaviour of the density in mGCG with these in the $f(R)$ and $\Lambda$CDM regimes:
\begin{equation}
	A = \left(3H_{\textrm{inf}}^2/\kappa^2\right)^{1+\lambda} ~~~~ B = \left(\rho_ra_0^4\right)^{1+\lambda}.
\end{equation}

In the later mGCG and $\Lambda$CDM eras, the correction term, $\Xi$, vanishes and $z''/z$ is reduced to the GR term which is determined as:
\begin{equation}
	\label{eq: a''a2}
	\frac{a''}{a}=\frac{\kappa^2}{6}a^2(\rho-3p).
\end{equation}
Here the energy density is defined by the second and third lines in the definition \eqref{eq: rhototal}, while the pressure is determined by the equations of state of the respective models: $p=(\rho-4A/\rho^{\lambda})/3$ for the mGCG; $p=\Sigma_i p_i = \Sigma w_i\rho_i$ for the $\Lambda$CDM model where $w=1/3$ for radiation, $w=0$ for dust and $w=-1$ for the cosmological constant.

Combining these results to write down \footnote{Notice that $z''/z=a''/a$ for $a>a_1$.} $z''/z$ during the three different eras we obtain:
\begin{align}
\label{eq: z''z}
\frac{z''}{z}=
\begin{cases}
	H_{\textrm{inf}}^2\left(2a^2-a_b^2\right) + \Xi &, a<a_1,\\
	\frac{2}{3}\kappa^2 a^2A\left[A +\frac{B}{a^{4(1+\lambda)}}\right]^{-\frac{\lambda}{1+\lambda}} &,~a_1<a<a_2,\\
	\frac{\kappa^2}{6}a^2\left[\rho_m \left(\frac{a_0}{a}\right)^3 + 4\rho_{\Lambda}\right] &,~a>a_2.
\end{cases}
\end{align}
The transition from the first to the second era takes place at $a=a_1$, and from the second to the third era at $a=a_2$. The values of both $a_1$ and $a_2$ are implicitly defined by the condition that the branches of Eqs~\eqref{eq: z''z} are continuous at the transition time.

\begin{figure}[t]
	\includegraphics[width=\columnwidth]{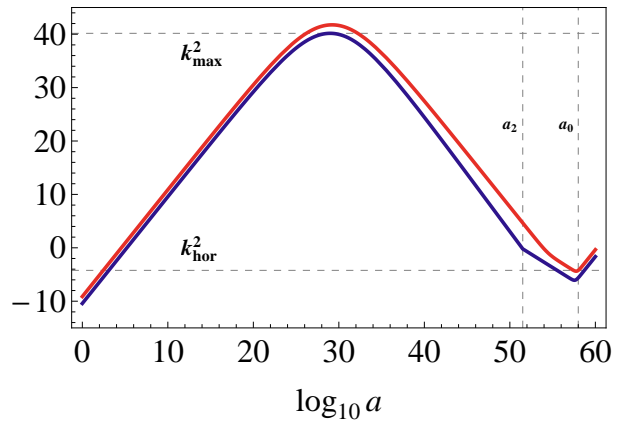}
	\caption[$z''/z$ and $a^2H^2$]{\label{fig: Potencial}This Figure shows: (i) the potential $a''/a$ (see the blue curve); (ii) the comoving wave-number $k_H^2=4\pi^2a^2H^2$ (see the red curve); as functions of the scale factor and plotted in a logarithmic scale in units of $m_p^2$.}
	
\end{figure}

To determine the gravitational energy spectrum $\Omega_{\textrm{GW}}$ we integrate equations \eqref{eq: X'eq} and \eqref{eq: Y'eqz} for a given wave-number $k$ from an initial time $t_{\textrm{ini}}$ set before the bounce until the present time. The wave-numbers $k$ are scanned from the mode that is entering the horizon at present, $k_{\textrm{hor}}=a_0H_0$, to the mode that corresponds to the maximum of $z''/z$, $k_{\textrm{max}}=\sqrt{\text{max}(z''/z)}$.
In order to understand the importance of the $f(R)$ corrections in equation \eqref{eq: Y'eqz}, we determine the evolution of $X$ and $Y$ using two methods. In the first method the integration is made considering solely the GR term $a''/a$ while in the second method the complete definition of $z''/z$ is used.

In order to solve the system\footnote{Notice that $z''/z$ reduces to $a''/a$ when $\Xi$ vanishes. Therefore, we can unify in Eqs.~\eqref{eq: X'eq} and \eqref{eq: Y'eqz}, the approaches we will follow in the GR and the $f(R)$ frameworks.} of differential equations, we need to set proper initial conditions. Even though the complex expression of the $f(R)$ term $z''/z$ gives little hope of obtaining an exact analytical solution, we can show that, when only the GR term, $a''/a$, or the approximation during the de Sitter-like regime are considered, the equations can be solved analytically. This allows us to obtain initial conditions for the numerical integration which we will apply on the next section.

We begin by combining the two first order differential equations, \eqref{eq: X'eq} and \eqref{eq: Y'eqz}, that govern the evolution of the variables $X$ and $Y$ in a single second order differential equation for X:
\begin{equation}
	\label{eq4: X''eq}
	X'' = \left(k^2 - \frac{z''}{z}\right)X.
\end{equation}
Also, we find that Eqs.~\eqref{eq: a''a1} and \eqref{eq: z''zLimit} can be cast in the general form $g_{\gamma}$:
\begin{equation}
	\label{eq: GenPot}
	g_{\gamma}(a)\equiv H_{\textrm{inf}}^2\left(2a^2-\gamma a_b^2\right),
\end{equation}
with $\gamma$ a parameter taking the value $\gamma=1$ for the GR case $(g_1=a''/a$, c.f. Eq.~\eqref{eq: a''a1}$)$  and $\gamma=2/3$ for the limiting de Sitter-like regime within the $f(R)$ framework $(g_{2/3}=z''/z(t\gg H_{\textrm{inf}}^{-1})$, c.f. Eq.~\eqref{eq: z''zLimit}$)$. Substituting $g_{\gamma}$ in \eqref{eq4: X''eq} and expressing the differential equation in terms of the cosmic time we obtain:
\begin{align}
	\label{eq: X''eqt}
	\ddot{X} +& H_{\textrm{inf}}\tanh(H_{\textrm{inf}}t)\dot{X}\nonumber\\
 &+ H_{\textrm{inf}}^2\left[2\cosh^2(H_{\textrm{inf}}t) - \gamma - q^2\right]X=0,
\end{align}
where the reduced wave-number $q$ is defined as $q\equiv k/(a_bH_{\textrm{inf}})$. Equation \eqref{eq: X''eqt} has a solution given by the linear combination:
\begin{equation}
	X(t) = D_1X_1(t) + D_2X_2(t),
\end{equation}
of the real-valued functions $X_1$ and $X_2$:
\begin{align}
	\label{eq: Xsolutions}
		X_1(t) &= \frac{\sqrt{\sinh^2\left(H_{\textrm{inf}}t\right) + \gamma + q^2}}{\gamma+q^2}
		\nonumber\\
		\times\cos&\left[\text{sgn}(\gamma-1+q^2)\frac{\phi_1(a)\sqrt{\gamma+q^2}-\phi_2(t)}{2}\right],
		\nonumber\\
		X_2(t) &= \frac{\sqrt{\sinh^2\left(H_{\textrm{inf}}t\right) + \gamma + q^2}}{\gamma+q^2}
		\nonumber\\
		\times\sin&\left[\text{sgn}(\gamma-1+q^2)\frac{\phi_1(a)\sqrt{\gamma+q^2}-\phi_2(t)}{2}\right].
	\end{align}
The phases $\phi_1$ and $\phi_2$ are defined as:
\begin{align}
	\label{eq: ThetaSolutions}
	\phi_1(t) &= \pi+ \text{sgn}(t)\arccos\left[2\text{sech}^2(H_{\textrm{inf}}t)-1\right],
	\nonumber\\
	\phi_2(t) &= \pi
	+ \text{sgn}(t)\arccos\left[\frac{2\left(\gamma+q^2\right)}{\sinh^2(H_{\textrm{inf}}t)+\gamma+q^2}-1\right],
\end{align}
The corresponding $Y$ functions can be obtained by inserting Eq.~\eqref{eq: Xsolutions} in Eq.~\eqref{eq: X'eq}:
\begin{align}
		\label{eq: Ysolutions}
		Y_1(t) &= \frac{i}{q}\left[\frac{\sinh(H_{\textrm{inf}}t)\cosh^2(H_{\textrm{inf}}t)}{\sinh^2(H_{\textrm{inf}}t)+\gamma+q^2}X_1(t) \right.
		\nonumber\\
		&- \left.\sqrt{\gamma+q^2}\frac{|\gamma-1+q^2|}{\sinh^2(H_{\textrm{inf}}t)+\gamma+q^2}X_2(t)\right],
		\nonumber\\
		Y_2(t) &= \frac{i}{q}\left[\frac{\sinh(H_{\textrm{inf}}t)\cosh^2(H_{\textrm{inf}}t)}{\sinh^2(H_{\textrm{inf}}t)+\gamma+q^2}X_2(t) \right.
		\nonumber\\
		&+ \left.\sqrt{\gamma+q^2}\frac{|\gamma-1+q^2|}{\sinh^2(H_{\textrm{inf}}t)+\gamma+q^2}X_1(t)\right].
\end{align}

The linear coefficients $D_1$ and $D_2$ are determined by the initial conditions $|X(t_{\textrm{ini}})|=|Y(t_{\textrm{ini}})|=1$, and by imposing that, for high $k$, $X$ converges to a Bunch-Davies-like solution \cite{Langlois:2010xc} with negative Hubble parameter, during the contraction phase.
\begin{align}
	D_1 = -i q\frac{\sqrt{X_4^2(t_{\textrm{ini}})+Y_2^2(t_{\textrm{ini}})}}{|\gamma-1+q^2|\sqrt{\gamma+q^2}},
\end{align}
\begin{align}
	D_2 &= \frac{-1}{\sqrt{X_4^2(t_{\textrm{ini}})+Y_2^2(t_{\textrm{ini}})}} \nonumber\\
	+  i&\frac{q}{|\gamma-1+q^2|\sqrt{\gamma+q^2}} \frac{X_3(t_{\textrm{ini}})X_4(t_{\textrm{ini}})+Y_1(t_{\textrm{ini}})Y_2(t_{\textrm{ini}})}{\sqrt{X_4^2(t_{\textrm{ini}})+Y_2^2(t_{\textrm{ini}})}}.
\end{align}

By fixing the initial conditions before the bounce, we are effectively defining a range of wave-numbers, $(k^2<z''/z(t_{\textrm{ini}}))$, that at the beginning of the integration are already outside the horizon. For those modes there is an increase in the graviton density, $|\beta|^2$, before the bounce. This fact is reflected in an upward shift in the energy spectrum, $\Omega_{\textrm{GW}}$, as we will show. As $z''/z$ grows exponentially with negative time, the further away from the bounce that we set the initial conditions, the more intense this shift is expected to be and the more wave-numbers are affected.

\section{\label{sec:Numerical}Numerical Simulations}

In our model we can identify a set of four independent parameters:
\begin{enumerate}
\item the time when the initial conditions are set, $t_{\textrm{ini}}$;
\item the scale factor at the time of the bounce, $a_b$;
\item the energy scale during the inflationary era, $E_{\textrm{inf}}$, which is related to the constant $H_{\textrm{inf}}$ by:
\begin{equation}
	E_{\textrm{inf}}=\left(\frac{3}{\kappa^2}H_{\textrm{inf}}^2\right)^{1/4};
\end{equation}
\item the value of the parameter $\lambda$ in the mGCG model. As this parameter changes the spectrum only in the high frequency range, as opposed to the effects of the bounce which are relevant only for the low frequencies, we use the value $\lambda = -1.04$ obtained in Ref.~\cite{Bouhmadi1} which corresponds to the best fit using the WMAP data \cite{WMAP}.
\end{enumerate}

For a chosen set of values $(a_b,t_{\textrm{ini}},E_{\textrm{inf}})$ we compute the GWs spectrum via the numerical integration of the differential equations \eqref{eq: X'eq} and \eqref{eq: Y'eqz} in two ways: case (i) within a GR perturbative treatment; case (ii) within a full $f(R)$ treatment. The integration is divided in two parts with the first one, during the $f(R)$ era, performed in terms of the cosmic time and the second one, during the mGCG and $\Lambda$CDM eras, performed in terms of the scale factor. Following previous works, \cite{Bouhmadi1,Mendes1,Sa1}, we stop the integration when $k^2\gg z''/z$, i.e. when the mode is well inside the horizon, since in this regime $X$ and $Y$ have a highly oscillatory sinusoidal behaviour while $|\beta|^2$ is virtually constant. This is an excellent way of reducing the computing time of the spectrum.

Before describing the two methods we define two particular wave-numbers. The first one:
\begin{equation}
	k_I=\sqrt{\max |\Xi|} =\sqrt{|\Xi_b|},
\end{equation}
can be interpreted as the smallest wave number for which the solutions in Eqs.~\eqref{eq: Xsolutions} and \eqref{eq: Ysolutions} (with $\gamma=2/3$) are a good approximation during the entire $f(R)$ era. The second one:
\begin{equation}
	\label{eq: kIIdef}
	k_{II}=\sqrt{z''/z(t_{\textrm{ini}})},
\end{equation}
indicates the mode with higher wave number that sees its graviton density increased during the contraction phase. Notice that, in principle, all modes should have an increase on its graviton number before the bounce and, indeed, if one sets $t_{\textrm{ini}}$ further away from the moment of the bounce the limit $k_{II}$ grows. However, by starting the integration from a given initial fixed time  for all the modes, only those modes close to the bounce will have a significant increases on its graviton density.

We proceed to describe the two methods employed to obtain the energy spectrum of the gravitational waves:

\subsection{GR approach for the perturbations}

In case (i) only the GR term $a''/a$ is considered when performing the integration, i.e. we integrate the Eqs.~\eqref{eq: X'eq} and \eqref{eq: Y'eq}. Since Eqs.~\eqref{eq: Xsolutions} and \eqref{eq: Ysolutions} $(\gamma=1)$ provide us with analytical solutions during the $f(R)$ era, we can skip the numerical integration all together during this period and use Eqs.~\eqref{eq: Xsolutions} and \eqref{eq: Ysolutions} to set the initial conditions for the integration during the mGCG era. This is done in two different ways depending on whether the mode has already crossed the horizon at the moment of the transition defined by $a_1$ or not (we remind that $a_1$ defines the transition from the $f(R)$ solution to the mGCG era (cf. Eq.\eqref{eq: z''z}).

 If the mode $k$ is outside of the horizon at the beginning of the mGCG era $(k\lesssim a_1H_1)$, we use Eqs.~\eqref{eq: Xsolutions} and \eqref{eq: Ysolutions} to fix the initial conditions for the integration at $a_i=a_1$ and then integrate numerically until  the mode reenters the horizon, i.e. until $a_f\sim10^2a_{\textrm{entry}}^{(k)}$, where $a_{\textrm{entry}}^{(k)}$ is the value of the scale factor at the moment of the reentry of the mode defined as $k=a_{\textrm{entry}}^{(k)}H(a_{\textrm{entry}}^{(k)})$.
As said previously, we stop our integration slightly after the mode reenters the horizon because, for $a > a_{\textrm{entry}}^{(k)}$, $X$ and $Y$ have an oscillatory behaviour; thus, $|\beta|$ remains constant after the crossing.

 For higher modes $(k\gtrsim a_1H_1)$, which are still inside the Hubble radius at the moment of the transition, we consider that the solutions \eqref{eq: Xsolutions} and \eqref{eq: Ysolutions} are valid slightly before the mode crosses the horizon, i.e., until $a_i\sim10^{-2}a_{\textrm{exit}}^{(k)}$ , where $a_{\textrm{exit}}^{(k)}$ is the value of the scale factor at the moment when the mode exits the horizon defined by $k=a_{\textrm{exit}}^{(k)}H(a_{\textrm{exit}}^{(k)})$. It is noteworthy to point out that, for large wave numbers, the solutions obtained for this model are virtually identical to the de Sitter solutions after the bounce and this is precisely the reason why we can apply them during the mGCG era when dominated by the effective cosmological constant $A$. Therefore, we use these solutions to set the initial conditions at $a=a_i$ and perform the integration until $a_f\sim10^2a_{\textrm{entry}}^{(k)}$, when the mode is again well inside the horizon. The reason we do not impose for the modes $a_1H_1\lesssim k$ the same boundary condition we used previously for $k\lesssim a_1H_1$ is to avoid the oscilatory regime before the modes cross the horizon as it is very time consuming.

\subsection{$f(R)$ approach for the perturbations}

In case (ii) the full definition of $z''/z$ is considered (see Eqs. \eqref{eq: zExp} and \eqref{eq: fRcorrec}), therefore we integrate Eqs.~\eqref{eq: X'eq} and \eqref{eq: Y'eqz}. The integration is performed accordingly to one of the following procedures:

If $(k\lesssim a_1H_1)$ and $(k\lesssim k_I)$, we begin the integration during the $f(R)$ era by setting the initial conditions at $t=t_{\textrm{ini}}$ using Eqs.~\eqref{eq: Xsolutions} and \eqref{eq: Ysolutions} $(\gamma=2/3)$. The system of differential equations is then integrated until the moment of the transition at $t=t_1$, where $t_1$ is defined by $a_1=a_b\cosh(H_{\textrm{inf}}t_1)$, as given in Eq.~\eqref{eq: scalefactor}. We then use these results as initial conditions for the integration during the mGCG era and then integrate until after the mode reenters the horizon at $a_f\sim10^2a_{\textrm{entry}}^{(k)}$.
 
Alternatively, if $(k\lesssim a_1H_1)$ and $(k\gtrsim k_I)$  the approximate solutions given by Eqs.~\eqref{eq: Xsolutions} and \eqref{eq: Ysolutions} $(\gamma=2/3)$ are not affected by the exact form of the potential $z''/z$ during the bounce. Therefore, we can skip the numerical integration during the $f(R)$ era, using those solutions to set the initial conditions for the integration during the mGCG era at $a_i=a_1$ and then we integrate numerically until the mode is well inside  the horizon at $a_f\sim10^2a_{\textrm{entry}}^{(k)}$.
 
If $(k\gtrsim a_1H_1)$ and $(k\lesssim k_I)$ we again set the initial conditions for the integration of the $f(R)$ period at $t=t_{\textrm{ini}}$ using Eqs.~\eqref{eq: Xsolutions} and \eqref{eq: Ysolutions} $(\gamma=2/3)$. The numerical integration is then performed in terms of the cosmological time until the mode approaches its horizon exit, i.e., $t=t_i$ defined by $a_i\equiv a_b\cosh(H_{\textrm{inf}}t_i)\sim10^{-2}a_{\textrm{exit}}^{(k)}$. These results are then used as initial conditions at $a=a_i$ and the integration is continued for the mGCG and $\Lambda$CDM eras  the mode are well inside the horizon at $a_f\sim10^2a_{\textrm{entry}}^{(k)}$.
 
Finally, if $(k\gtrsim a_1H_1)$ and $(k\gtrsim k_I)$ we can use the approximated solutions Eqs.~\eqref{eq: Xsolutions} and \eqref{eq: Ysolutions} (with $\gamma=2/3$) to set the initial conditions for the integration of the mGCG era before the mode exits the horizon at $a_i\sim10^{-2}a_{\textrm{exit}}^{(k)}$. The integration is then performed until $a_f\sim10^2a_{\textrm{entry}}^{(k)}$, when the mode is again well inside the horizon.

\subsection{Numerical results}

The range of frequencies scanned is delimited by $\omega_{\textrm{min}}= k_{\textrm{hor}}/a_0\approx 1.43\times 10^{-17}$ rad s$^{-1}$ and $\omega_{\textrm{max}}= k_{\textrm{max}}/a_0\approx 1.49\times 10^{-11}E_{\textrm{inf}}$ rad s$^{-1}$. The values of the parameters of the $\Lambda$CDM model were taken from the  WMAP7 data \cite{WMAP}: $\Omega_r=8\times 10^{-5}$, $\Omega_m=0.272$, $\Omega_{\Lambda}=0.728$ and $H_0=$70.4 km s$^{-1}$Mpc$^{-1}$. The value of scale factor at present time is set to $a_0=10^{58}$.

Finally the following constraints on the energy spectrum of GWs were considered \cite{Sa1,Smith1}:
\begin{itemize}
\item From the CMB radiation: $h_0^2\Omega_{\textrm{GW}}(\omega_{\textrm{hor}},\eta_0)\lesssim 7\times 10^{-11}$ for $h_0=H_0/(100$ km s$^{-1}$Mpc$^{-1}$) and $\omega_{\textrm{hor}}=2\times 10^{-17}h_0$ rad s$^{-1}$;
\item From observation of milliseconds pulsar: $h_0^2\Omega_{\textrm{GW}}(\omega_{\textrm{pul}},\eta_0)< 2\times 10^{-8}$ for $\omega_{\textrm{pul}}=2.5\times 10^{-8}h_0$ rad s$^{-1}$;
\item From the Cassini spacecraft: $h_0^2\Omega_{\textrm{GW}}(\omega_{\textrm{Cas}},\eta_0)< 0.014$ for $\omega_{\textrm{Cas}}=7.5\times 10^{-6}h_0$ rad s$^{-1}$;
\item From the LIGO experiment: $h_0^2\Omega_{\textrm{GW}}(\omega_{\textrm{LIGO}},\eta_0)< 3.4\times 10^{-5}$ for frequencies on the order of a few hundred rad s$^{-1}$;
\item From BBN: $h_0^2\Omega(\omega,\eta_0)d\omega/\omega < 5.6\times 10^{-6}$ for $\omega_n \approx 10^{-9}$rad s$^{-1}$.
\end{itemize}

\begin{figure*}[t]
	\centering
	\subfloat[]{\includegraphics[width=.48\textwidth]{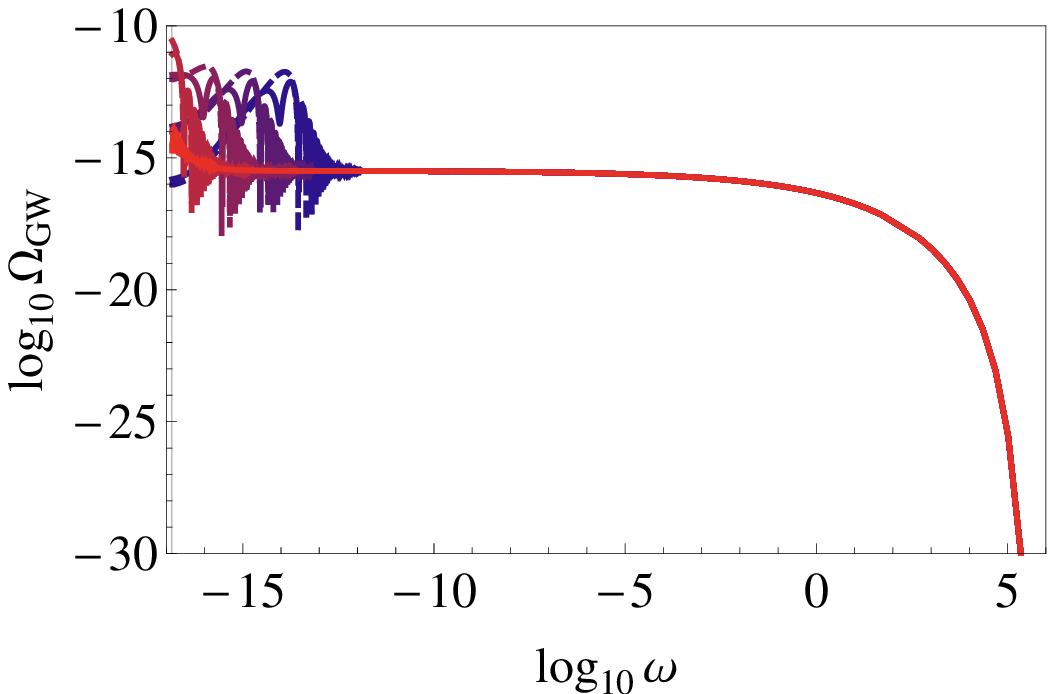}}
	\hfill
	\subfloat[]{\includegraphics[width=.48\textwidth]{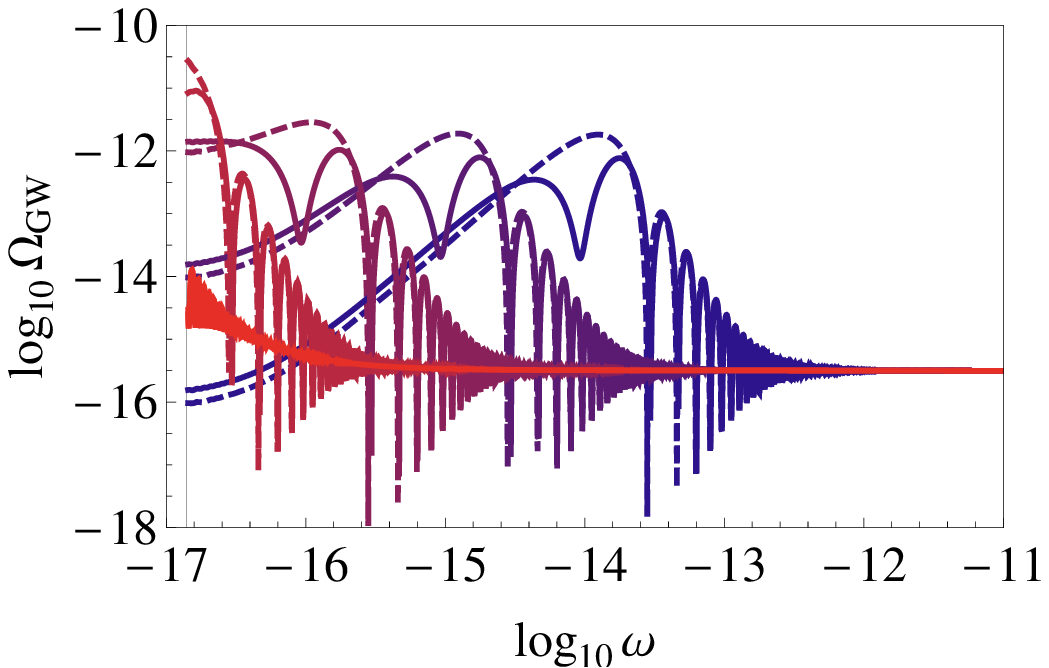}}
	\caption[Energy spectrum of GWs for varying $a_b$]{\label{fig: SpectraAi}Spectra of the energy density of GWs at present time for different values of the scale factor at the bounce, $a_b$. The spectra are plotted for \textbf{(a)} the entire range of frequencies and \textbf{(b)} the low frequency range. A comparison is made between the results obtained using the GR setup (dashed curves) and the $f(R)$ treatment (continuous curves). The value of $a_b$  increases from the red curve to the blue curve: $a_b=2\times 10^2$; $a_b=2\times 10^3$; $a_b=2\times 10^4$; $a_b=10\times 10^5$; $a_b=10\times 10^6$. $(E_{\textrm{inf}}=1.5\times 10^{16}$GeV; $a_{\textrm{ini}}=10a_b)$}
\end{figure*}

\begin{figure*}[t]
	\centering
	\subfloat[]{\includegraphics[width=.48\textwidth]{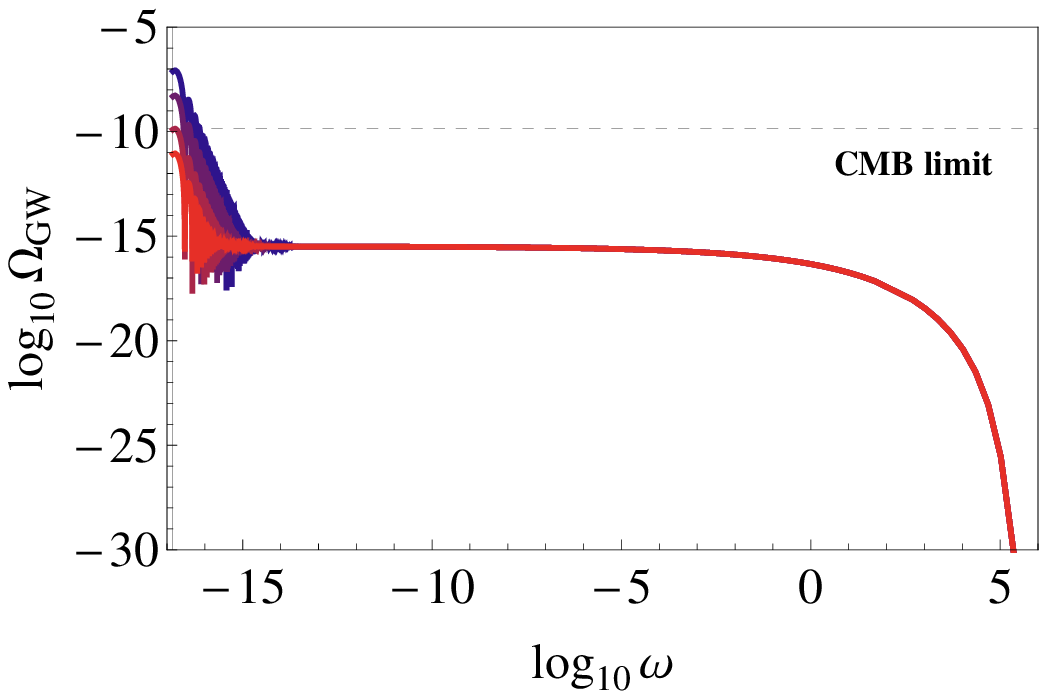}}
	\hfill
	\subfloat[]{\includegraphics[width=.48\textwidth]{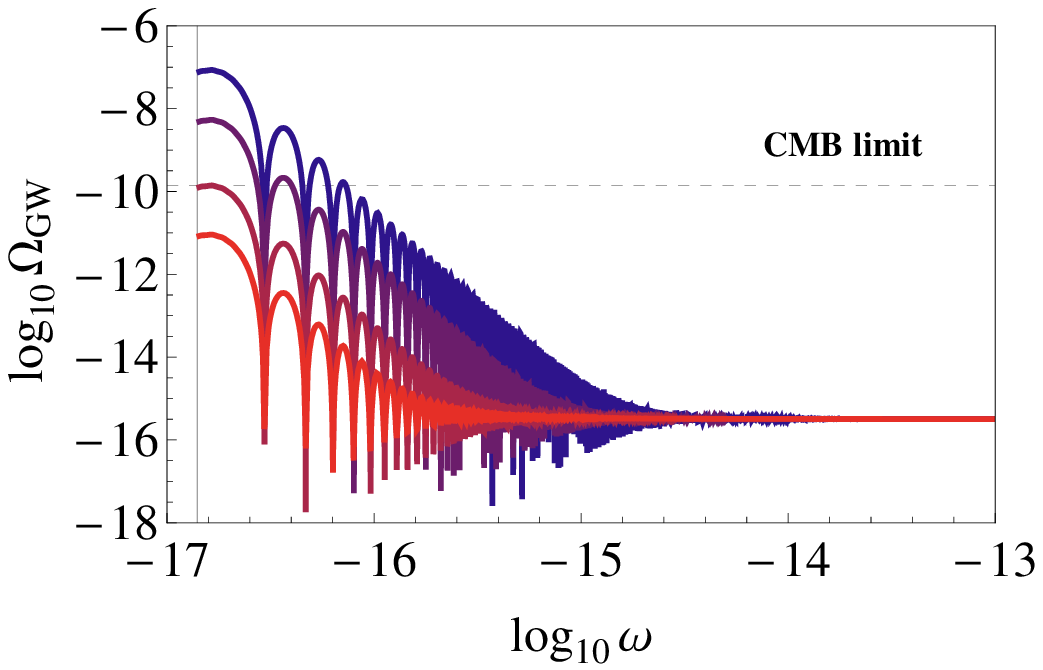}}
	\caption[Energy spectrum of GWs for varying $a_{\textrm{ini}}/a_b$]{\label{fig: SpectraAini}Spectra of the energy density of GWs at present time for different values of the initial time, $t_{\textrm{ini}}$, which is related to the amount of contraction prior to the bounce by $H_{\textrm{inf}}t_{\textrm{ini}}=\textrm{arccosh}(a_{\textrm{ini}}/a_b)$. The spectra are plotted for \textbf{(a)} the entire range of frequencies and \textbf{(b)} the low frequency range. The value of $a_{\textrm{ini}}/a_b$  increases from the red curve to the blue curve: $a_{\textrm{ini}}=10a_b$; $a_{\textrm{ini}}=20a_b$; $a_{\textrm{ini}}=50a_b$; $a_{\textrm{ini}}=100a_b$. $(E_{\textrm{inf}}=1.5\times 10^{16}$GeV; $a_b=2000)$.}
\end{figure*}

\begin{figure*}[t]
	\centering
	\subfloat[]{\includegraphics[width=.48\textwidth]{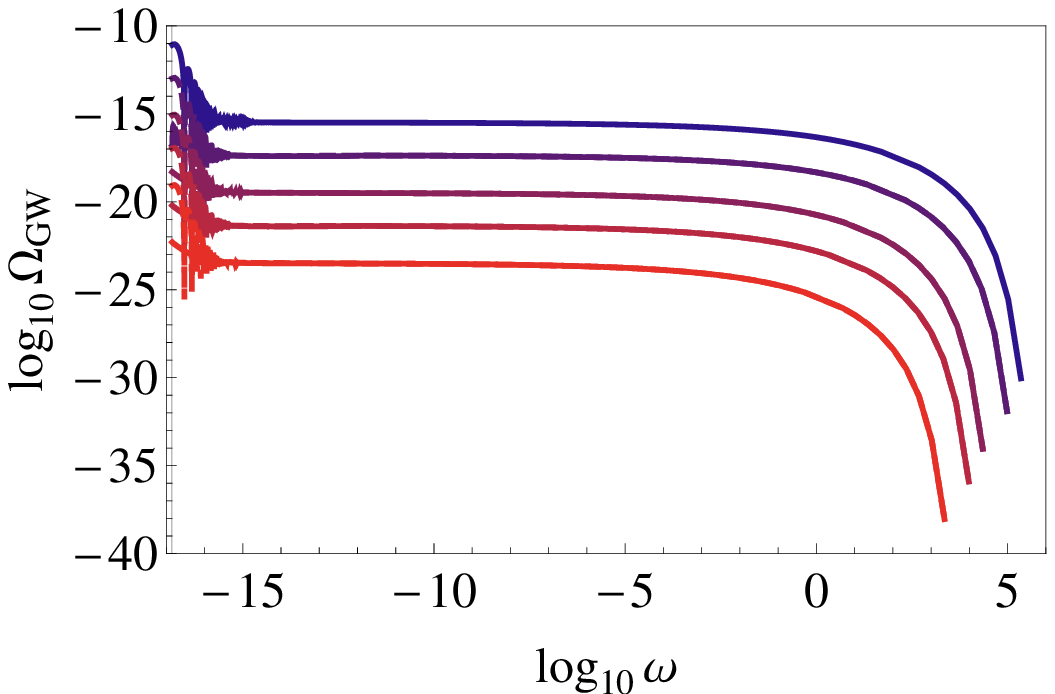}}
	\hfill
	\subfloat[]{\includegraphics[width=.48\textwidth]{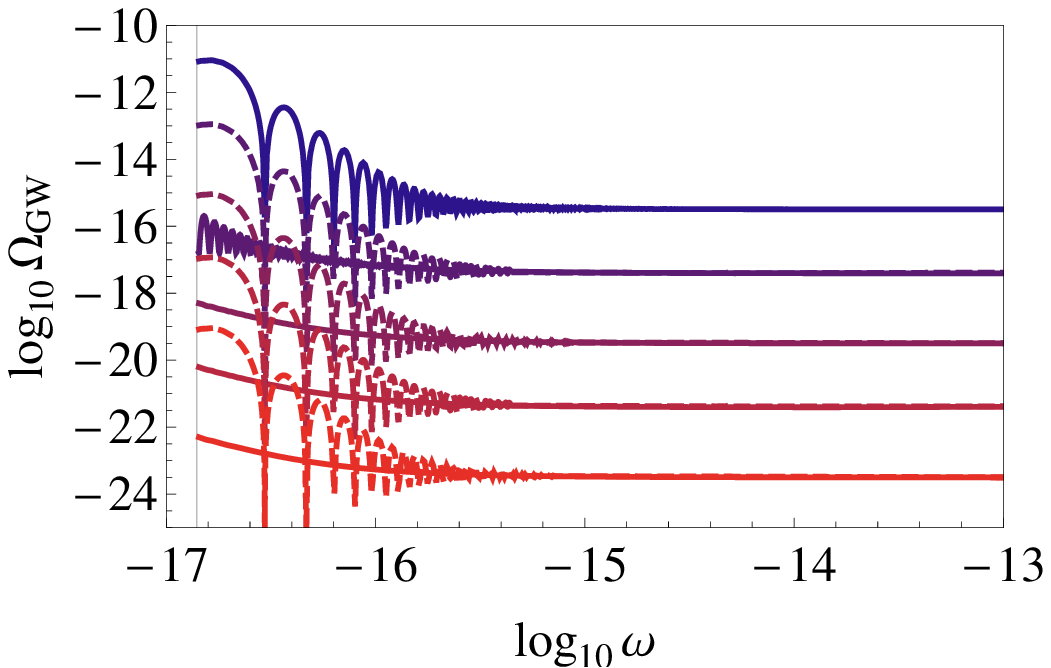}}
	\caption[Energy spectrum of GWs for varying $E_{\textrm{inf}}$]{\label{fig: SpectraEi}Spectra of the energy density of GWs at present time for different values of the energy scale during inflation, $E_{\textrm{inf}}$. The spectra are plotted for \textbf{(a)} the entire range of frequencies and \textbf{(b)} the low frequency range. A comparison is made between the results obtained with a fixed value of $a_b$ (continuous curves) and a fixed value of $a_bH_{\textrm{ini}}$ (discontinuous curves). The value of $E_{\textrm{inf}}$  increases from the red curve to the blue curve: $E_{\textrm{inf}}=1.5\times 10^{14}$GeV; $E_{\textrm{inf}}=0.5\times 10^{15}$GeV; $E_{\textrm{inf}}=1.5\times 10^{15}$GeV; $E_{\textrm{inf}}=0.5\times 10^{16}$GeV; $E_{\textrm{inf}}=1.5\times 10^{16}$GeV. $(a_{\textrm{ini}}=10a_b$; $a_b=2\times 10^3)$.}
\end{figure*}

The results obtained for the GWs spectra are shown in Figs.~\ref{fig: SpectraAi}, \ref{fig: SpectraAini} and \ref{fig: SpectraEi}. In each figure two of the three parameters $(a_{\textrm{ini}}/a_b,a_b,E_{\textrm{inf}})$ are fixed while the third one is changed so as to illustrate its effects on the spectrum. The energy scale for the numerical integration is chosen within the allowed observational  values and such that the effect of the bounce is enhanced on the spectrum of the GWs. The initial time, $t_{\textrm{ini}}$, is fixed before the bounce, during the phase of de Sitter-like contraction, and the value of the scale factor at the bounce is chosen so that the effect on the low frequency of the spectrum are amplified. In Fig.~\ref{fig: SpectraAi} the results obtained using the perturbative approach in (i) GR and (ii) $f(R)$ are also compared.

As expected, the existence of a bounce in the early universe affects the spectrum only in the low frequency range, where a highly oscillatory regime is present in contrast with the smooth plateau in the intermediate frequencies and the rapid decay in the high frequency range. The fact that the oscillatory structure appears in the spectra of both treatments (i) and (ii) suggests that it is due to the existence of the bounce and not a consequence of the effects of $f(R)$-gravity. Similar oscillations have been obtained in works of loop quantum cosmology first pointed out by Afonso et al \cite{Afonso1,Sa2}, and in the spectrum of scalar perturbation of non-singular models with scalar fields \cite{Piao:2003zm}.

The differences in the spectra obtained in (i) a GR setup and in (ii) an $f(R)$ approach are evidentiated  in Fig.~\ref{fig: SpectraAi}. These appear only for very low values of the reduced wave-number, $q$, which corresponds to the low frequencies of the spectra where an extra local minimum occurs for $k=a_bH_{\textrm{inf}}/\sqrt{3}$ . This corresponds to the value that makes the factor $\text{sgn}(\gamma-1+q^2)=0$ in solutions \eqref{eq: Xsolutions} and \eqref{eq: Ysolutions} for $\gamma=2/3$. Notice that for the medium and high frequencies, and even for the higher frequencies on the oscillatory regime, the spectra obtained with the two treatments are virtually indistinguishable. The spectra obtained are in conformity with the assumption that the maximum value of the $f(R)$ corrections, $\Xi$, sets the scale on the upper limit of the wave-numbers for which there are noticeable differences in the spectra determined using the methods (i) or (ii). Thus, for the gravitational spectrum $\Omega_{\textrm{GW}}$ to be useful in discriminating between the GR and $f(R)$ theories describing the evolution of the tensor perturbations, the wave-number $k_I$ must be within the accessible range of wave-numbers today., i.e.:
\begin{equation}
	\label{eq: DiffCond}
	k_I > k_{\textrm{hor}}.
\end{equation}

In Fig.~\ref{fig: SpectraAi} a horizontal displacement is observed in the spectrum when the value of $a_b$ is modified. This reflects the fact that the solutions \eqref{eq: Xsolutions} and \eqref{eq: Ysolutions} do not depend directly on the wave-number $k$ but instead on the reduced wave-number $q=k/(a_bH_{\textrm{inf}})$. As such, it is expected that the position of the peaks on the spectrum to be affected the by changes on the product $a_bH_{\textrm{inf}}$. This effect is also seen in Fig.~\ref{fig: SpectraEi}, because lowering the energy scale for a fixed value of $a_bH_{\textrm{inf}}$ only shifts the spectrum downwards without changing the position of the peaks.

In Fig.~\ref{fig: SpectraAini} an increase on the energy spectrum of GWs, $\Omega_{\textrm{GW}}$, as well as in the range of the affected frequencies, is observed as the initial conditions are set further away from the bounce (higher $a_{\textrm{ini}}/a_b$). This results agree with the analysis done previously in Sec.~\ref{sec:Spectrum}.

The CMB constraints set an upper limit in the low frequency end of the spectrum of $\Omega_{\textrm{GW}}(\omega_{\textrm{hor}},\eta_0)\approx1.4\times 10^{-10}$. This value limits the amount of contraction that one takes into account as the intensity of the peaks in the energy spectrum scales rapidly with increasing $a_{\textrm{ini}}/a_b$. In fact from Fig.~\ref{fig: SpectraAini} it is possible to observe that if the condition \eqref{eq: DiffCond} is met the relation $a_{\textrm{ini}}/a_b$ is heavily constrained: for $a_{\textrm{ini}}/a_b=20$ the maximum of the potential is slightly above $10^{-10}$, while for $a_{\textrm{ini}}/a_b=50$ the maximum of the potential is almost two orders of magnitude higher which clearly violates the CMB limit.
\begin{figure}[t]
	\includegraphics[width=\columnwidth]{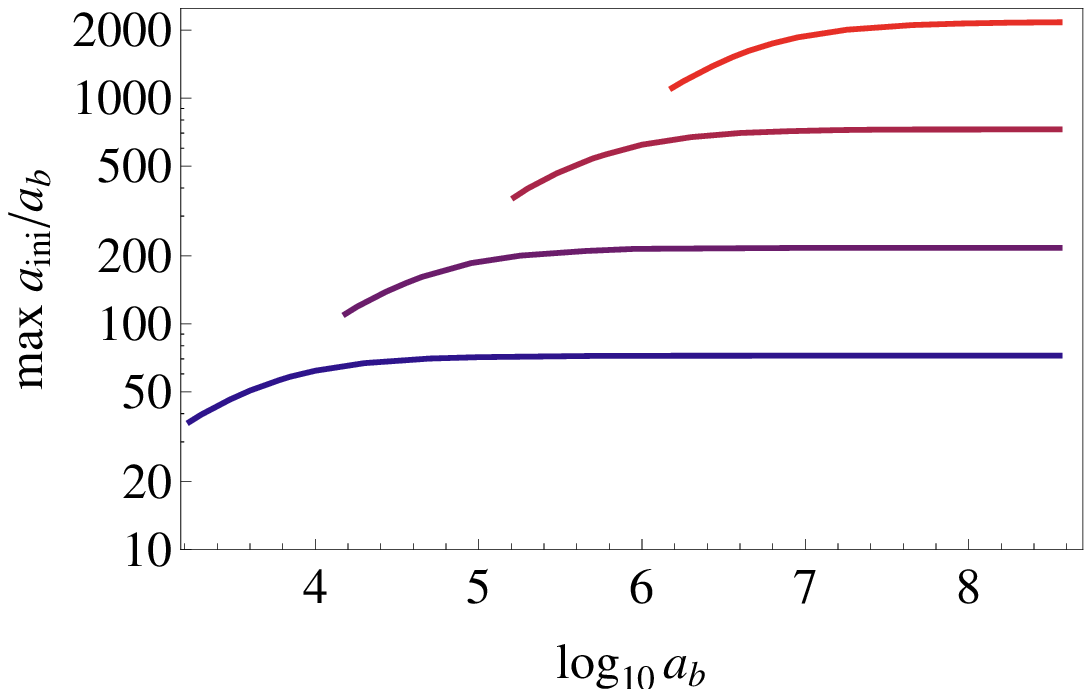}
	\caption[]{\label{fig: t0max} This picture shows the maximum allowed value of $a_{\textrm{ini}}/a_b$ that does not violate the CMB limit, as a function of $a_b$ and for different values of the energy scale, $E_{\textrm{inf}}$. From the red curve to the blue curve: $E_{\textrm{inf}}=0.5\times 10^{15}$GeV; $E_{\textrm{inf}}=1.5\times 10^{15}$GeV; $E_{\textrm{inf}}=0.5\times 10^{16}$GeV; $E_{\textrm{inf}}=1.5\times 10^{16}$GeV. ($k=a_bH_{\textrm{inf}}$)}
\end{figure}
In order to obtain a better picture of the restrictions set on $a_{\textrm{ini}}/a_b$, we select the wave-number of the most leftward peak (of the order of $k_p\approx a_bH_{\textrm{inf}}$) and repeat the integration with varying $a_{\textrm{ini}}/a_b$. The results obtained are presented in Fig.~\ref{fig: t0max} for different inflationary energy scales. While the maximum allowed value of $a_{\textrm{ini}}/a_b$ increases slightly with $a_b$, it rapidly becomes stagnant with growing $a_b$. We call such parameter, for a fixed energy scale, of $l_{E_{\textrm{inf}}}(a_b)$. The curves $l_{E_{\textrm{inf}}}(a_b)$ transform with a rescaling of the energy scale, $E_{\textrm{inf}}\rightarrow mE_{\textrm{inf}}$, as:
\begin{align}
	\label{limit_contraction}
	l_{E_{\textrm{inf}}}\left(a_b\right)\sim m^{-1}l_{mE_{\textrm{inf}}}\left(m^{-2}a_b\right),
\end{align}
Here, $m$ is a dimensionless scaling factor. From Eq.~\eqref{limit_contraction} we see that by lowering the energy scale, the limiting value of the relation $a_{\textrm{ini}}/a_b$ is shifted upwards and the minimum value of $a_b$ for which $k_p>k_{\textrm{hor}}$  increases. For $E_{\textrm{inf}}=1.5\times10^{16}$GeV we obtain $l_{E_{\textrm{inf}}}(a_b)\approx 72$ in the high $a_b$ region (see the asymptotic behaviour of Fig.~\ref{fig: t0max}. Notice that this constrain in the amount of the considered contraction can be avoided by lowering the product $a_bH_{\textrm{inf}}$, so that the peaks that violate the CMB limit are shifted toward the left on the spectrum, to the region of frequencies not accessible today. This, however, means that one would lose the ability to differentiate between the GR and the purely $f(R)$ imprints on the spectrum, or, in the extreme case, lose the oscillatory structure altogether.

In Fig.~\ref{fig: SpectraEi} the effects of changing the energy density during inflation may be observed. Besides the scaling of $k_{\textrm{max}}$ with $E_{\textrm{inf}}$ ($k_{\textrm{max}}\propto E_{\textrm{inf}}$), a vertical shift of the intermediate plateau is observed, with:
\begin{align}
	\frac{\Omega_{\textrm{GW}}(E_{\textrm{inf}}^{(1)})}{\Omega_{\textrm{GW}}(E_{\textrm{inf}}^{(2)})}\sim \left(\frac{E_{\textrm{inf}}^{(1)}}{E_{\textrm{inf}}^{(2)}}\right)^4.
\end{align}
As discussed above, the position of the peaks observed on the spectrum depend primarily on the relation $a_bH_{\textrm{inf}}$, which in turn goes as $E_{\textrm{inf}}^2$. Therefore, changing the value of the energy scale during inflation is bound to impose a horizontal displacement of the peaks, as observed in  Fig.~\ref{fig: SpectraEi}. For comparison, we also plot the spectra obtained by changing the value of $E_{\textrm{inf}}$ while maintaining the value of $a_bH_{\textrm{inf}}$ constant. In this case the peaks do not change their position but the spectrum is displaced  vertically.


\section{\label{sec:Conclusions}Conclusions}

In this work we investigate the effects that the presence of a bounce in the early universe has on the energy spectrum of the gravitational waves. In order to avoid violations of the null energy condition that usually appear in bouncing FLRW cosmologies in GR \cite{Molina1}  we work within $f(R)$ gravity \cite{Capozziello2} fixing the desired behaviour for the scale factor and then solving the Friedmann equations to obtain the function $f$ (\textit{designer} $f(R)$ gravity \cite{Dunsby:2010wg,Carloni:2010ph}). A mGCG model \cite{Bouhmadi1} was used to connect the bouncing and subsequent inflationary era with the radiation epoch as it provides an easy way to describe a smooth transition between the two phases (inflation and radiation) without affecting the low frequencies of the spectrum.

The method of the Bogoliubov coefficients \cite{Parker1} was used to determine the evolution of the graviton density while treating the tensorial perturbations both in a GR approach and in a full $f(R)$ treatment \cite{Capozziello2}. The energy spectrum of the gravitational waves was determined for different values of the scale factors at the time of the bounce, different amounts of contraction before the bounce and for different energy scales during the inflationary era.

We show that the presence of the bounce leads to a higher creation rate of low frequency gravitons, which results in an oscillatory signature in the low frequency range of the spectrum, with various peaks whose position and intensity depend on the relations $a_bH_{\textrm{inf}}$ and $a_{\textrm{ini}}/a_b$. Here, $a_b$ defines the value of the scale factor at the time of the bounce, $H_{\textrm{inf}}$ is related to the energy scale during inflation by $H_{\textrm{inf}}^2={\kappa^2}/{3}\;E_{\textrm{inf}}^4$ and $a_{\textrm{ini}}$ defines the value of the scale factor at a fixed time before the bounce when the density of gravitons is negligible. An increase of  $a_bH_{\textrm{inf}}$ shifts the peaks rightward on the spectrum while an increase in $a_{\textrm{ini}}/a_b$ results in an enhancement of the peaks that correspond to wave-numbers below $k_{II}$ (cf. Eq.~\eqref{eq: kIIdef}). The intermediate and high frequency regions are left unaffected by the bounce. The condition \eqref{eq: DiffCond}, which we have assumed, ensures that the modes currently entering the horizon have some imprints of the $f(R)$ era.

The results obtained are within all the observational constraints considered except for the CMB radiation data as, for the higher energy scales even a contraction of two orders of magnitude before the bounce makes the spectrum violate the limits imposed on the low frequencies of the spectrum \cite{Sa1,Smith1}. An estimation on the maximum amount of the allowed  contraction without violating the CMB constraints was determined by studying the dependence of the height of the most intense peak of the spectrum on the parameters of the model. For an energy scale of $E_{\textrm{inf}}=1.5\times10^{16}$GeV during the early inflation, we obtained $\max(a_{\textrm{ini}}/a_b)\lesssim72$. The fact that the contraction before the bounce his highly constrained may be viewed as an indication of the existence of an extra state of the universe that precedes the de Sitter-like contraction. The potential imprint of such a hypothetical state on the gravitational waves spectrum is not studied in this work.

Furthermore, we note that the results obtained in this work are hard to be detected in the near future. Most of the planned  experiments that might be sensitive to the cosmological gravitational waves, namely BBO \cite{BBO} and DECIGO \cite{DECIGO} operate on frequencies of $\gtrsim 10^{-5}$Hz (see Fig. 2 of Ref.~\cite{Smith1} and Fig. 6 of Ref.~\cite{Kawasaki:2012rw}), while the imprints originated from the bounce are observable, at best, up to frequencies of $\lesssim 10^{-9}$Hz. The best hope for the detection of these effects relies in future measurement of the B-mode polarization of the CMB radiation \cite{Bourhrous:2012kr,Efstathiou:2009xv,Bonaldi:2011vv}.

Finally, we point out that the solution obtained for $f(R)$ is differentiable for all the values of $R$ but its second and higher derivatives blow up at the bounce. Although this divergence does not have serious implications in the tensorial perturbations, it does imply that the scalaron \cite{Starobinsky:2007hu} becomes a tachyon near the bounce, so some problems are expected when studying the scalar sector. A way to overcome this is to consider further corrections on the gravitational action that have been proved to be free from such kind of instabilities \cite{Biswas:2012bp}.

\acknowledgments

The authors are grateful to S. Capozziello, P. Dunsby and G. Olmo for very useful comments on a previous version of the manuscript and to the anonymous referees for their constructive feedbacks.
M.B.L. is supported by the Spanish Agency ``Consejo Superior de Investigaciones Cient\'{\i}ficas" through JAEDOC064 and the Basque Foundation for Science IKERBASQUE. This work was supported by the Portuguese Agency ``Funda\c{c}\~{a}o para a Ci\^{e}ncia e Tecnologia" through PTDC/FIS/111032/2009.

\end{document}